\documentclass{article}
\usepackage{amssymb}
\usepackage{amsmath}
\usepackage{amsthm}

\newtheorem{prpstn}{Proposition}
\newtheorem{thrm}[prpstn]{Theorem}
\newcommand{\xR}{\mathbb{R}}

\begin{document}

\title{Acyclic Orientations with Path Constraints}

\author{Rosa~M.~V.~Figueiredo\\
Instituto de Matem\'atica e Estat\'\i stica, UERJ\\
20550-900 Rio de Janeiro - RJ, Brazil\\
\\
Valmir~C.~Barbosa\thanks{Corresponding author (valmir@cos.ufrj.br).}
\quad
Nelson~Maculan\\
Programa de Engenharia de Sistemas e Computa\c c\~ao, UFRJ\\
Caixa Postal 68511, 21941-972 Rio de Janeiro - RJ, Brazil\\
\\
Cid~C.~de Souza\\
Instituto de Computa\c c\~ao, Unicamp\\
Caixa Postal 6176, 13084-971 Campinas - SP, Brazil}

\date{}

\maketitle

\begin{abstract}
Many well-known combinatorial optimization problems can be stated over the set of acyclic orientations
of an undirected graph. For example, acyclic orientations with certain diameter constraints are
closely related to the optimal solutions of the vertex coloring and frequency assignment problems. 
In this paper we introduce a linear programming formulation of acyclic orientations
with path constraints, and discuss its use in the solution of the vertex coloring problem and
some versions of the frequency assignment problem. A study of the polytope associated with the 
formulation is presented, including proofs of which constraints of the formulation are facet-defining
and the introduction of new classes of valid inequalities.

\end{abstract}

\section{Introduction}
\label{sec:intro}

Let $G=(V,E)$ be an undirected graph, $V$ its set of vertices, and $E$ its set of edges. 
An orientation of $G$ is a function $\omega$ with domain $E$ such that
$\omega([i,j])\in \{(i,j),(j,i)\}$ for all $[i,j]\in E$. That is, $\omega$ assigns a direction to each
edge in $E$. Given an orientation $\omega$, let $G_{\omega}$ be the directed
graph obtained by replacing each edge $[i,j]\in E$ with the arc $\omega([i,j])$. An orientation
is said to be acyclic if $G_{\omega}$ contains no directed cycle. 

Many combinatorial optimization problems can be solved by determining an optimal orientation
of a graph with respect to some measure of optimality. 
One example is the linear ordering problem \cite{Junger85B},
also called the permutation problem or the triangulation problem, 
and closely related to the acyclic subdigraph problem \cite{Junger85}.
Given a complete directed graph $D_n=(V_n,A_n)$ on $n$ vertices and arc weights $c_{ij}$ for each arc 
$(i,j)\in A_n$, the linear ordering problem consists in finding a spanning acyclic tournament in $D_n$
such that the sum of the weights of its arcs is as large as possible. The spanning acyclic tournament
in $D_n$ is equivalent to an acyclic orientation of the complete undirected graph with $n$ vertices.

Another problem whose solution is given by an optimal orientation is discussed 
in \cite{Bermond99}. Given an undirected graph $G$ and an orientation $\omega$, not necessarily acyclic,
the distance $\textit{dist}(i,j)$ from a vertex $i$ to another vertex $j$ is the length of the shortest path
from $i$ to $j$ in $G_{\omega}$. The diameter of $G_{\omega}$ is defined in \cite{Bermond99} 
as $\max_{i,j\in V} dist(i,j)$.
An annular network is a graph that can be represented as a two-dimensional grid consisting of a number of
concentric circles around a center and some straight lines crossing all the circles. The problem of finding 
an orientation of minimum diameter when $G$ is an annular network is useful in various applications \cite{Bermond99}.

A communications network can be modeled as the undirected graph $G$ if vertices represent processors
and edges communications links between pairs of processors. Each vertex in $V$ must be assigned a set of buffers 
in order to store messages which move through the network. For every network there exists a lower
bound on the number of buffers which have to be maintained at each vertex to allow deadlock-free routing.
In \cite{bermond97acyclic} an approach to prevent deadlocks is investigated in which finding the minimum 
number of buffers is related to finding an optimal acyclic orientation $\omega$ of $G$. The optimality 
criterion used is this case is to minimize the maximum number of changes of orientations on some directed 
paths in $G_{\omega}$.

Knowledge about acyclic orientations can also be useful in the solution of the vertex coloring problem.
Roy \cite{Roy67} e Gallai \cite{Gallai68}, independently, showed that, given an undirected graph $G$,
the length of a longest elementary path of each possible orientation of $G$ yields an upper bound on 
the chromatic number of $G$. Also, they proved that the exact value of the chromatic number is accomplished 
by the minimum bound over the set of all orientations of $G$. In a subsequent work \cite{Deming79},
Deming showed that it is sufficient to consider just the set of acyclic orientations of $G$. 
In the same work, Deming also showed that using another measure of optimality, again over the set of the acyclic 
orientations of $G$, it is possible to find the maximum independent set of that graph. In this case,
and for $\omega$ an acyclic orientation,
the criterion is to maximize the size of a minimum chain decomposition of $G_{\omega}$.   

Since frequency assignment problems are closely related to the vertex coloring problem,
it would be expected that an orientation-based approach could also be proposed for that problem.
This is done in \cite{Borndorfer98}, where a two-stage integer programming model is proposed in which
the outer stage consists of an acyclic subdigraph problem with additional longest-path constraints.
Also, in \cite{Maniezzo2000} an orientation model is pointed at as a promising way to obtain
good lower bounds to be incorporated into a metaheuristic approach to
the frequency assignment problem.

All these applications have motivated us to the study of the acyclic orientations of an undirected 
graph. In the present study, we introduce an integer programming formulation of the set of 
all acyclic orientations with constraints defined over a set of paths. 
As we demonstrate in Section~\ref{sec:combProblems}, there are well-known combinatorial optimization 
problems that relate clearly to acyclic orientations with path constraints.

We close this section by giving the necessary definitions and notations to be used throughout the paper. 
Let $G=(V,E)$ be an undirected graph. A coloring of $G$ is an 
assignment of labels to each vertex such that the end-vertices of any edge have different labels. The chromatic 
number of $G$ is the smallest number of different labels needed to define a coloring of $G$ and is denoted 
by $\chi(G)$. The vertex coloring problem is the problem of providing $G$ with a coloring that employs 
$\chi(G)$ labels. Let $D=(V,A)$ be a digraph. 
If $D$ is an acyclic digraph, the diameter of $D$ with respect to a vector $c\in\xR^{|A|}$ of arc weights 
is the length of a longest weighted path in $D$.
Given an arc set $B\subseteq A$, we denote by $D[B]$ the subdigraph of $D$ 
induced by $B$. If $D[B]$ is an acyclic digraph, then we say that $B$ is an acyclic arc set of $D$.
For the sake of conciseness, we henceforth use path to refer to an elementary path and equate a path
with its arc set.

The remainder of the paper is structured as follows. In Section~\ref{sec:combProblems} we relate 
the vertex coloring and frequency assignment problems to orientations with diameter constraints. 
A model of acyclic orientations with another kind of path constraints is presented
in Section~\ref{sec:formulation}. We prove that this model can also be used to find an
orientation with diameter constraints. In Section~\ref{sec:polytope} we investigate 
the polyhedral structure of the polytope associated with the model proposed.
Finally, in Section~\ref{sec:directions} we discuss directions for further investigation.

\section{Combinatorial optimization problems and\\
acyclic orientations with diameter constraints}
\label{sec:combProblems}

As noted above, in \cite{Roy67,Gallai68} the chromatic number $\chi(G)$ of an undirected
graph $G$ is described as an optimization problem over the set of all orientations of
$G$. This result is revised in \cite{Deming79}, where the author proves that it is sufficient
to consider only the set of acyclic orientations, that is,
\begin{equation}
\chi(G)=1+\min_{\omega\in\Omega}\max_{p\in {\mathcal P}_{\omega}} |p|,
\label{teo:Deming}
\end{equation}
\noindent where $\Omega$ denotes the set of all acyclic orientations of $G$, ${\mathcal P}_{\omega}$ 
the set of paths in $G_{\omega}$, and $|p|$ the number of arcs in $p$. 
Alternatively, (\ref{teo:Deming}) asks for $\omega$ such that $G_{\omega}$ has the least possible
diameter considering arc weights equal to 1, which is then $\chi(G)-1$.

Let $\omega^*$ be an acyclic orientation that solves (\ref{teo:Deming}). The source decomposition 
of the graph $G_{\omega^*}$ gives us, in polynomial time, a coloring of $G$ that uses exactly $\chi(G)$ labels. 
An acyclic directed graph has a unique source decomposition defined as a partition of the vertex set
$V$ into sets $V_1,V_2,\ldots,V_k$, where $V_i$ is the set of all vertices of the subgraph induced
by $V\setminus\{V_1\cup\cdots\cup V_{i-1}\}$ whose in-degree is zero. Notice that each vertex set
$V_i$ in the partition $(V_1,V_2,\ldots,V_k)$ is an independent set, and then the partition
defines a coloring of $G$. Then, in order to solve the vertex coloring problem for a graph $G$ it is sufficient to 
find an acyclic orientation $G_{\omega^*}$ with minimum diameter.
Henceforth, we let $q(G)=\chi(G)-1$ denote the length of the minimum diameter over all acyclic
orientations of $G$.

In the literature, different problems can be found under the common heading of frequency assignment
problem (FAP). In all of them, the classic approach relates the problem to the vertex coloring problem.
It then follows from our previous discussion that we can consider using acyclic orientations also to 
solve this problem. 
This is done in \cite{Borndorfer98}, where a new orientation model is proposed for a version of FAP, 
and in \cite{Maniezzo2000}, where the LP bounds of an orientation model are suggested to be 
incorporated into a metaheuristic proposed for another version of FAP. In order to describe the idea 
of an orientation model for FAP, we next introduce the elements usually found in the definition of that 
problem. Let $\mathcal{L}$ denote an index set of links, $\mathcal{F}_i$ a set of available 
frequencies for link $i\in \mathcal{L}$, and $d_{ij}$ a channel separation that defines 
the minimum distance between frequencies assigned to links $i,j\in \mathcal{L}$. The inexistence 
of a channel separation between links $i$ and $j$ can be imposed by setting $d_{ij}=0$. 
A frequency assignment specifies, for each $i\in \mathcal{L}$, a frequency in $\mathcal{F}_i$ 
for link $i$.

The most common version of FAP asks for an assignment of frequencies that minimizes the number of 
frequencies used, known as the frequency spectrum, while satisfying the channel separation imposed by 
$d_{ij}$ for all $i,j\in \mathcal{L}$. Clearly, when every $d_{ij}$ is equal 
to zero or one and, for each $i\in \mathcal{L}$, $F_i=\mathcal{L}$, then FAP can be easily
cast into the vertex coloring problem and can therefore be solved by looking for an acyclic 
orientation of minimum diameter. 
In other FAP versions,
given a fixed frequency spectrum $\Phi$ we must find an assignment of frequencies to the links 
that minimizes a cost function, sometimes an interference function, defined over the channel 
separation constraints. Next we describe the orientation model presented in \cite{Maniezzo2000} 
for one of these versions of FAP.

We use a graph $G=(V,E)$ with $V$ being the set of links $\mathcal{L}$. 
A pair of vertices $i,j\in V$ is connected by an edge $[i,j]\in E$ if and only if there is a 
distance requirement imposed by a channel separation $d_{ij}>0$ 
on the frequencies that are to be assigned to $i$ and $j$. Let $f_i$ be a positive integer 
variable specifying the frequency assigned to link $i$, $\zeta_{ij}$ a binary variable which
is equal to 1 if and only if the channel separation between $f_i$ and $f_j$
is not guaranteed, and $c_{ij}$ the cost of violating the channel separation constraint.
In order to specify an orientation of  edge $[i,j]\in E$, a binary decision variable $o_{ij}$ 
is introduced and defined as follows:
$o_{ij}=1$ if edge $[i,j]\in E$ is oriented from $i$ to $j$ and $o_{ij}=0$ 
if edge $[i,j]\in E$ is oriented from $j$ to $i$. The orientation
model is as follows:
\begin{align}
\mathrm{minimize\ }    & \textstyle\sum_{[i,j]\in E} c_{ij}\zeta_{ij}                      &\nonumber\\
\mathrm{subject\ to\ } & f_j - f_i + \Phi\zeta_{ij} \geq d_{ij}o_{ij} - M(1-o_{ij}), &\forall\ [i,j]\in E \label{oij=1}\\
                     & f_i - f_j + \Phi\zeta_{ij} \geq d_{ij}(1-o_{ij}) -Mo_{ij}, &\forall\ [i,j]\in E \label{oij=0} \\
                     & f_i\in \mathcal{F}_i,                               &\forall\ i\in V\nonumber\\
                     & o_{ij},\zeta_{ij}\in\{0,1\}.               &\forall\ [i,j]\in E\nonumber
\end{align}
In this formulation, $M$ is an arbitrary, large constant that makes either (\ref{oij=1})  
or (\ref{oij=0}) active in case $o_{ij}$ is equal to 1 or 0, respectively. These constraints 
try to impose the channel separation required for $[i,j]\in E$. 
In the solution of this orientation model, $o_{ij}=1$ means
that the frequency assigned to link $i$ will be smaller than
that assigned to link $j$, unless $\zeta_{ij}$ assumes value 1. Likewise, $o_{ij}=0$
means that the frequency assigned to link $i$ has to be greater than that assigned to link $j$,
again unless $\zeta_{ij}$ assumes value 1. Thus, in this orientation model the
variables $o_{ij}$, $[i,j]\in E$, define the set of all orientations of graph $G$.

Notice that we can reformulate the orientation model as follows. For each edge $[i,j]\in E$, let us define a binary 
variable $o_{ji}$, besides variable $o_{ij}$, such that $o_{ij}=1$ if edge $[i,j]$ is oriented 
from $i$ to $j$ and $o_{ji}=1$ if edge $[i,j]$ is oriented from $j$ to $i$.
According to the new set of orientation variables, we rewrite (\ref{oij=1}) 
and (\ref{oij=0}) as
\begin{align}
&f_j - f_i \geq d_{ij}o_{ij} - M(1-o_{ij}),&\forall\ [i,j]\in E\nonumber\\
&f_i - f_j \geq d_{ij}o_{ji} - M(1-o_{ji}).&\forall\ [i,j]\in E\nonumber
\end{align}
Now an orientation of an edge cannot always be defined, but when this happens the channel separation
is imposed by one of the new constraints. In order for the objective function to retain its
meaning, we must require $\zeta_{ij}=1-o_{ij}-o_{ji}$. 
It is not difficult to see that a solution to this alternative model 
is given by an acyclic orientation of a subset of $E$. Additionally,
the solution is given by an acyclic orientation of 
a subset of $E$ with diameter at most $\Phi$ with respect to arc weights $d_{ij}=d_{ji}$.

The orientation model presented in \cite{Borndorfer98} is very similar to the model discussed above,
but with another definition for channel separations and, consequently, another definition for the objective
function. The approach presented by the authors is based on a two-stage integer programming model 
in which the outer stage solution is related to an orientation of $G$ with diameter constraints. 
FAP instances are described with channel separation equal to 1, 2, and 3, 
but no computational result is presented. They mention the need for a linear integer formulation 
for acyclic orientations of diameter at most $\Phi$ with respect to arc weights 
given by channel separations, and point out some interesting questions to be answered on 
the structure of the associated polyhedron.

So far we have described how some important, but difficult, combinatorial optimization problems relate to
the acyclic orientations of a graph with diameter constraints. The remainder of the paper
is devoted to studying a formulation of acyclic orientations under binary
arc weights. While this can be seen to be immediately useful to the case of vertex coloring and
of FAP instances with channel separation equal to 1, there are no 
real FAP instances of this type. But channel separations equal to 
2 or 3 can also be handled, since we can construct an equivalent,
enlarged instance in which all channel separations
are 1. Suppose vertices $i$ and $j$ have channel separation $d_{ij}=2$. We define a new vertex $v_{ij}$,
new edges $[i,v_{ij}],[j,v_{ij}]$ with channel separation equal to 1, and eliminate the edge $[i,j]$. 
Additionally to the constraints of the 
orientation model, it clearly suffices that we introduce the following new constraints:
$o_{iv_{ij}}+o_{jv_{ij}}\leq 1$ and $o_{v_{ij}i}+o_{v_{ij}j}\leq 1$.
An analogous transformation can be done in the cases in which $d_{ij}=3$.

In the next section we enunciate a formulation of acyclic orientations with other path constraints 
used to describe acyclic orientations with diameter constraints. 
In the sequel, we refer to the diameter under arc weights that equal 1 simply as diameter.

\section{Acyclic orientations with path constraints}
\label{sec:formulation}

Let $G=(V,E)$ be an undirected graph with $n=|V|$ vertices and $m=|E|$ edges,
and $\kappa$ a positive scalar. We want to describe the acyclic orientations $\omega$ 
of $G$ that minimize a measure defined over the set of all paths in 
$G_{\omega}$ with $\kappa$ arcs. Let $D=(V,A)$ be a directed graph with 
$A=\{(i,j),(j,i)\mid\forall\ [i,j]\in E\}$. For each arc $(i,j)\in A$, we introduce a 
binary decision variable $w_{ij}$ such that
$w_{ij} = 1$ and $w_{ji} = 0$ if $\omega([i,j])=(i,j)$, or
$w_{ij} = 0$ and $w_{ji} = 1$ if $\omega([i,j])=(j,i)$.
Let $P_\kappa(A)$ denote the set of all paths in $D$ with $\kappa$ arcs.
Likewise, let $C(A)$ be the set of all cycles in $D$.
The formulation follows:
\begin{align}
\mathrm{minimize\ }   & z                                 &\label{fo1}\\
\mathrm{subject\ to\ }& w_{ij} + w_{ji} = 1,               &\forall\ [i,j]\in E \label{des1}\\ 
                    &\textstyle\sum_{(i,j)\in C} w_{ij} \leq |C|-1,&\forall\ C \in C(A) \label{des2}\\ 
                    &\textstyle\sum_{(i,j)\in p} w_{ij}\leq z,    &\forall\ p \in P_\kappa(A)\label{des3}\\
                    & w_{ij} \in \{0,1\},                &\forall\ (i,j) \in A \label{des4}\\
                    & 0 \leq z \leq \kappa.             &\label{des5}       
\end{align}
The constraints in (\ref{des1}), (\ref{des2}), and (\ref{des4}) define the acyclic orientations 
of $G$. Once we have a vector $w\in\xR^{2m}$ satisfying these constraints, let $G_w=(V,A_w)$ be 
the digraph obtained by directing the edges in $E$ according to
the variables $w_{ij}$, i.e., a digraph with vertex set $V$ and arc set $A_w$ defined
as: $(i,j)\in A_w$ if and only if $w_{ij}=1$. The variable $z$ is a continuous variable that establishes 
an upper bound on the overall number of arcs oriented in the same direction in any path of $D$ with at most $\kappa$
arcs. The constraints in (\ref{des3}) and (\ref{des5}) give this meaning to variable $z$.
Finally, the objective function in (\ref{fo1}) makes variable $z$ assume the minimum possible upper bound.
Let us refer to this formulation as $AO(G,\kappa)$.

The following propositions will guide us in the search for an acyclic orientation with minimum diameter
using the above formulation. Recall that $q(G)=\chi(G)-1$.

\begin{prpstn}
If $(\bar{w},\bar{z})$ is a feasible solution to $AO(G,\kappa)$ with $\kappa\leq q(G)$, then
$\bar{z}=\kappa$.
\label{prop:z=k}
\end{prpstn}

\begin{proof}
Let $(\bar{w},\bar{z})$ be a feasible solution of the formulation. 
Let us assume that $\bar{z} < \kappa = q(G)$. From (\ref{des3}) 
we conclude that the length of each path in $G_{\bar{w}}$ is less than $q(G)$, i.e.,
the diameter of $G_{\bar{w}}$ is less than $q(G)$, and we get a contradiction.
Thus, $\bar{z}\geq \kappa$. The result follows, since $\bar{z}\leq\kappa$. 
\end{proof}

\begin{prpstn}
If $(w^*,z^*)$ is an optimal solution to $AO(G,\kappa)$ with $\kappa\geq q(G)+1$, then
$z^*<\kappa$.
\label{prop:z<k}
\end{prpstn}

\begin{proof}
Consider an acyclic orientation of $G$ with diameter equal to $q(G)$, and let $(\bar{w},\bar{z})$ be a solution that
induces such an orientation. Let $p$ be a path in $D$ with $\kappa$ arcs. Since $\kappa\geq q(G)+1$ and 
the diameter of $G_{\bar{w}}$ is $q(G)$, in each succession of $q(G)+1$ arcs in $p$ there is at least one arc 
with $w_{ij}=0$, thus $(\bar{w},\bar{z})$ with $\bar{z}=\kappa-\lfloor\kappa/(q(G)+1)\rfloor<\kappa $ 
is a feasible solution of $AO(G,\kappa)$.
The result follows, since the objective function of $AO(G,\kappa)$ minimizes the value of $z$.
\end{proof}

Now let $\mathit{UB}$ be an upper bound on the minimum diameter $q(G)$ of $G$ and solve $AO(G,\kappa)$
with $\kappa=\mathit{UB}$. 
If $(w^*,z^*)$ is the optimal solution found and $z^*=\kappa$, then from Proposition~\ref{prop:z<k}
it follows that $q(G)=\kappa$. If $z^*<\kappa$, then by Proposition~\ref{prop:z=k} we can reduce the parameter
$\kappa$ by finding the diameter of $G_{w^*}$ and assigning its value to $\kappa$. 
Repeating the entire procedure at most $\mathit{UB}$ times clearly yields an acyclic orientation of 
$G$ with minimum diameter. 

As discussed in Section~\ref{sec:combProblems}, the solution of some FAP variants can be found by looking for an
acyclic orientation with diameter equal to at most a given frequency spectrum $\Phi$.
Notice that, if we consider $\kappa=\Phi+1$ and fix $z=\Phi$, then the constraints of $AO(G,\kappa)$ 
describe the acyclic orientations with diameter at most $\Phi$.

\section{On the acyclic subgraph with path constraints polytope}
\label{sec:polytope}

The acyclic orientation with path constraints polytope, defined as the convex hull of all feasible solutions 
of the model described in Section~\ref{sec:formulation}, is not a full-dimensional one. 
Let us then consider the following alternative formulation, which models an acyclic subgraph
with path constraints:
\begin{align}
\mathrm{minimize\ }   &\textstyle z - (m+1)\sum_{(i,j)\in A} w_{ij}     &\nonumber\\
\mathrm{subject\ to\ }& w_{ij} + w_{ji} \leq 1,            &\forall\ [i,j]\in E \label{des6}\\
                    &\textstyle\sum_{(i,j)\in C} w_{ij} \leq |C|-1,&\forall\ C \in C(A) \label{des7}\\
                    &\textstyle\sum_{(i,j)\in p} w_{ij}\leq z,    &\forall\ p \in P_\kappa(A) \label{des8}\\
                    & w_{ij} \in \{0,1\},                &\forall\ (i,j) \in A \label{des9}\\
                    & 0\leq z \leq \kappa.              &\label{des10}
\end{align}

Notice that (\ref{des6}) allows us to not define an orientation of the edges in $E$.
This is where the alternative objective function comes in, since it penalizes any solution
that does not orient an edge.
It is also clear that all feasible solutions of $AO(G,\kappa)$ are contained in the set of feasible solutions
of this new formulation. Moreover, it is not difficult to verify that, if $(w^*,z^*)$ is
an optimal solution of this formulation, then $w^*$ defines an acyclic orientation of $G$
and $z^*$ is the smallest value of $z$ satisfying (\ref{des8}). Let us refer to this acyclic 
subgraph model with path constraints as $AS(G,\kappa)$.

The polytope $P_{G,\kappa}$ associated with $AS(G,\kappa)$ is defined as
\[P_{G,\kappa}=\mathrm{conv}\{ (w,z)\in \xR^{2m}\times\xR\mid
(w,z)\mbox{ satisfies (\ref{des6})--(\ref{des10})}\}.\] 
Now we turn our attention to the structure of $P_{G,\kappa}$, where we recall that $\kappa>0$.
Given an acyclic arc set $B$ of $D$ (i.e., $B\subseteq A$), the incidence vector $w^B\in\xR^{2m}$ of 
$B$ is defined as follows:
$w^B_{ij}=1$, if $(i,j)\in B$, and $w^B_{ij}=0$, if $(i,j)\not\in B$. Next, the dimension of $P_{G,\kappa}$ 
is established.

\begin{thrm}
The polytope $P_{G,\kappa}$ is full-dimensional, i.e., $\mathrm{dim}(P_{G,\kappa}) = 2m + 1$.
\label{kappa:dimensao}
\end{thrm}

\begin{proof}
Since $P_{G,\kappa}$ contains the null vector, it is sufficient to present other $2m+1$ linearly independent
solutions $(w,z)\in \xR^{2m}\times\xR$ in $P_{G,\kappa}$. For each arc $(i,j)\in A$, let 
$B_{ij}=\{(i,j)\}$. The $2m$ solutions $(w^{B_{ij}},1)$, together with the solution $(w,1)$ with $w$
being the null vector, are clearly linearly independent.
\end{proof}

The following theorems establish which constraints of $AS(G,\kappa)$ define facets of $P_{G,\kappa}$.

\begin{thrm}\emph{[Trivial inequalities]}
\begin{itemize}
\item[(a)] For all $(i,j)\in A$, $w_{ij} \geq 0$ defines a facet of $P_{G,\kappa}$;
\item[(b)] for all $(i,j)\in A$, $w_{ij} \leq 1$ is not a facet-defining inequality for $P_{G,\kappa}$;
\item[(c)] the inequality $z \geq 0$ is not a facet-defining inequality for $P_{G,\kappa}$;
\item[(d)] the inequality $z \leq \kappa$ defines a facet of $P_{G,\kappa}$.
\end{itemize}
\label{kappa:triviais}
\end{thrm}

\begin{proof} The proof is straightforward.
\end{proof}

Notice now that, in $AS(G,\kappa)$, the constraints in (\ref{des6}) can be seen as equivalent to the
constraints in (\ref{des7}) with $|C|=2$. We then have the following.

\begin{thrm}\emph{[Cycle inequality]}
Let $C\in C(A)$. The inequality
\[\sum_{(i,j)\in C} w_{ij} \leq |C|-1\]
defines a facet of $P_{G,\kappa}$ if and only if $|C|\leq \kappa$.
\label{kappa:ciclo}
\end{thrm}

\begin{proof}
Let $F=\{(w,z)\in P_{G,\kappa}\mid\sum_{(i,j)\in C}  w_{ij} = |C|-1\}$ be the face of $AS(G,\kappa)$ 
defined by the cycle inequality written for the cycle $C$. We assume that there is an inequality 
$a^Tw + bz \leq c$ valid for $P_{G,\kappa}$ such that 
$F\subseteq F_{ab}= \{(w,z)\in P_{G,\kappa} \mid a^Tw+bz=c\}$
and show that the inequality defining $F_{ab}$ can be written as a positive scalar multiple of the cycle 
inequality defining $F$. 
Let $\{v_1,v_2,\ldots,v_{|C|}\}$ be the vertex set of cycle $C$ and $p_{v_k\rightarrow v_l}$
the path from vertex $v_k$ to vertex $v_l$ on $C$. Also, let us assume $v_{|C|+1}=v_1$. 
Consider an arc $(r,v)\in A\setminus C$. Define an acyclic arc set $B_1$ as $B_1=p_{v_k\rightarrow v_{k-1}}$
if $r=v_k\in V(C)$ and $v=v_l\in V(C)$, or $B_1= p_{v_1\rightarrow v_{|C|}}$ in the other cases. Also, define another
acyclic arc set $B_2=B_1\cup \{(r,v)\}$. From the solutions $(w^{B_1},\kappa)$ and $(w^{B_2},\kappa)$ 
in $a^Tw+bz=c$ we can conclude that $a_{rv}=0$.
Now consider the acyclic arc set $B_3$ defined as $B_3= p_{v_1\rightarrow v_{|C|}}$. The solutions
$(w^{B_3},\kappa)$ and $(w^{B_3},\kappa-1)$ lead to $b=0$.
Now we prove the relations among the non-null coefficients in $a^Tw+bz\leq c$.
Suppose $a_{v_1v_2}=\gamma$. Consider the arc $(v_2,v_3)$ and define the arc sets $B_4=p_{v_2\rightarrow v_1}$
and $B_5=p_{v_3\rightarrow v_2}$. The solutions defined by these sets and $z=\kappa-1$ imply $a_{v_2v_3}=\gamma$.
Repeating this argument yields $a_{rv}=\gamma$ for all $(r,v)\in C$. 
Finally, the cycle inequality defined by a cycle with $|C|>\kappa$ is not a facet-defining one, since 
in this case constraint $z\leq \kappa$ dominates the cycle inequality.
\end{proof}

The methodology to be used in all facet-defining proofs will be the same as in the proof of Theorem~\ref{kappa:ciclo}.
The following theorem establishes the necessary and sufficient conditions for each inequality in (\ref{des8}) to be
facet-defining for $P_{G,\kappa}$.

\begin{thrm}\emph{[Path inequality]}
Let $p\in P_\kappa(A)$ and let $s$ and $t$ be,
respectively, the source and sink of $p$. The inequality
\[\sum_{(i,j)\in p} w_{ij} - z \leq 0\]
defines a facet of $P_{G,\kappa}$ if and only if $[s,t]\not\in E$.
\label{kappa:caminho}
\end{thrm}

\begin{proof}
Let us assume that $\{v_1,v_2,\ldots,v_{\kappa+1}\}$ is the set of vertices defining $p$.
Consider an arc $(r,v)\in A\setminus p$ and define an arc set $B_1$ as follows:
(i) if $r=v_k$, $k\geq 2$, and $v\not\in \{v_1,v_2,\ldots,v_{\kappa+1}\}$: $B_1=p\setminus \{(v_{k-1},v_k)\}$;
(ii) if $r\not\in \{v_1,v_2,\ldots,v_{\kappa+1}\}$ and $v=v_k$, $k\leq \kappa$:  $B_1=p\setminus \{(v_k,v_{k+1})\}$;
(iii) if $r=v_k$ and $v=v_l$, $k>l$: $B_1=p\setminus \{(v_{k-1},v_k)\}$;
(iv) in any other case: $B_1=p\setminus \{(v_1,v_2)\}$. Also, define an arc set
$B_2=B_1\cup \{(r,v)\}$. From the solutions $(w^{B_1},\kappa-1)$ 
and $(w^{B_2},\kappa-1)$ we can conclude that $a_{rv}=0$.
Now assume $a_{v_1v_2}=\gamma$ and define the arc sets $B_3=p\setminus \{(v_1,v_2)\}$ and
$B_4=p\setminus \{(v_2,v_3)\}$. The solutions $(w^{B_3},\kappa-1)$ and $(w^{B_4},\kappa-1)$ 
allow us to conclude that $a_{v_2v_3}=a_{v_1v_2}=\gamma$. Repeating the same argument we 
obtain $a_{rv}=\gamma$ for every arc $(r,v)\in p$. 
Now consider the arc sets $B_5$ and $B_6$ defined as $B_5=p\setminus \{(v_1,v_2)\}$ and $B_6=p$.
From the solutions $(w^{B_5},\kappa-1)$ and $(w^{B_6},\kappa)$, we conclude that $b=-a_{v_1v_2}=-\gamma$.
Finally, we argue that path constrains cannot be facet-defining when $[s,t]\in E$.
Clearly, when $[s,t]\in E$ all solutions satisfying the path inequality with equality also satisfy
$w_{ts} = 0$. 
\end{proof}

Next we present new valid inequalities for $P_{G,\kappa}$ which are related to
some substructures of $D$. While a cycle inequality is induced by any cycle in $D$,
the following theorem introduces another valid inequality induced by cycles with 
$\kappa+1$ arcs.

\begin{thrm}\emph{[Cycle-$z$ inequality]}
Let $C\in C(A)$ be such that $|C|=\kappa+1$. The inequality
\[\sum_{(i,j)\in C} w_{ij} \leq z\]
defines a facet of $P_{G,\kappa}$.
\label{teo:cicloZ}
\end{thrm}

\begin{proof}
The validity proof is trivial due to (\ref{des7}). 
The facet-defining proof uses the notation introduced in Theorem~\ref{kappa:ciclo}. 
Also, using arguments analogous to the ones in the proof of Theorem~\ref{kappa:ciclo} we 
can conclude that $a_{rv}=0$ for every arc $(r,v)\in A\setminus C$ and that 
$a_{rv}=\gamma$ for every arc $(r,v)\in C$. Now consider the arc sets $B_1$ and $B_2$ 
defined as $B_1=p_{v_1\rightarrow v_{|C|}}$ and $B_2=B_1\setminus\{(v_1,v_2)\}$.
From the solutions $(w^{B_1},\kappa)$ and $(w^{B_2},\kappa-1)$, we conclude that $b=-a_{v_1v_2}=-\gamma$.
\end{proof}

Our next theorem establishes a class of valid inequalities induced by paths with length equal 
to $\kappa-1$ whose vertices have a common adjacent vertex outside the path.

\begin{thrm}\emph{[Path-$(\kappa-1)$ inequality]}
Let $p$ be a path in $D$ with $\kappa-1$ arcs and $\{v_1,v_2,\ldots,v_{\kappa}\}$ its vertex set. 
If a vertex $u\in V\setminus\{v_1,v_2,\ldots,v_{\kappa}\}$ exists such that $[v_k,u]\in E$
for all $k\in \{1,2,\ldots,\kappa\}$, then the inequality
\[\sum_{(i,j)\in p} w_{ij} + \sum_{k\in\{1,\ldots,\kappa\}} (w_{uv_k}+w_{v_ku}) - \kappa + 1 \leq z\]
is valid for $P_{G,\kappa}$.
\label{teo:caminhoK-1}
\end{thrm}

\begin{proof}
Let $(\bar{w},\bar{z})$ be a feasible solution in $P_{G,\kappa}$.
If $\sum_{k\in\{1,\cdots,\kappa\}} (\bar{w}_{uv_k}+\bar{w}_{v_ku}) < \kappa$, then the inequality is
trivially satisfied. Assuming this is not the case, the result follows from noticing that, for any
feasible orientation of the edges $[v_k,u]\in E$ with $k\in \{1,2,\ldots,\kappa\}$, a suitable path inequality
exists involving vertices $u$ and $v_1,\ldots,v_\kappa$.
\end{proof}

Paths in $D$ with $\kappa-2$ arcs also induce valid inequalities for $P_{G,\kappa}$, as the 
next theorem demonstrates.

\begin{thrm}\emph{[Path-$(\kappa-2)$ inequality]}
Let $p$ be a path in $D$ with $\kappa-2$ arcs and $\{v_1,v_2,\ldots,v_{\kappa-1}\}$
its vertex set. If vertices $u,r\in V\setminus\{v_1,v_2,\ldots,v_{\kappa-1}\}$ 
exist such that $[v_1,u],[v_{\kappa-1},u]\in E$ and $[r,u]\in E$, then the inequality
\[\sum_{(i,j)\in p} w_{ij} + w_{ur} + w_{ru} \leq z\]
is valid for $P_{G,\kappa}$.
\label{kappa:caminhoK-2}
\end{thrm}

\begin{proof}
Let  $(\bar{w},\bar{z})$ be a feasible solution in $P_{G,\kappa}$.
If $\bar{w}_{ur} + \bar{w}_{ru} < 1$, then the validity of the inequality follows trivially. 
Assuming this is not the case, the result follows from applying (\ref{des8}) 
to $p$ extended by $\{(v_{\kappa-1},u),(u,r)\}$ if $\bar{w}_{ur}=1$, and to $p$ preceded by
$\{(r,u),(u,v_1)\}$ if $\bar{w}_{ru}=1$.
\end{proof}

A cycle in $D$ having length $\kappa$ gives rise to yet another class of valid inequalities for 
$P_{G,\kappa}$.

\begin{thrm}\emph{[Cycle-arcs inequality]}
Let $C\in C(A)$ be such that $|C|=\kappa$ and 
$\{v_1,v_2,\ldots,v_{\kappa}\}$ be its vertex set. Suppose in addition that $\{r_1,r_2,\ldots,r_{\kappa}\}\subset V$ 
and $A'=\{e_1,e_2,\dots,e_{\kappa}\}\subset A$ exist such that:
\begin{itemize}
\item [(i)] $\{r_1,r_2,\ldots,r_{\kappa}\}\cap \{v_1,v_2,\ldots,v_{\kappa}\}=\emptyset$,
\item [(ii)] either $e_k=(r_k,v_k)$ for all $1\leq k\leq \kappa$ or $e_k=(v_k,r_k)$ for all $1\leq k\leq \kappa$. 
\end{itemize}
Then the inequality
\[\sum_{(i,j)\in C} (\lfloor{\kappa/2}\rfloor w_{ij} + w_{ji}) +  \sum_{(i,j)\in A'} w_{ij} 
- \kappa \leq \lfloor{\kappa/2}\rfloor z \]
is valid for $P_{G,\kappa}$.
\label{teo:cicloArcos}
\end{thrm}

\begin{proof}
Consider the cycle-arcs inequality rewritten as 
\begin{equation}
\lfloor{\kappa/2}\rfloor(z - \sum_{(i,j)\in C} w_{ij}) + \kappa - \sum_{(i,j)\in A'} w_{ij} 
\geq \sum_{(i,j)\in C} w_{ji}.
\label{des2:cicloArcos}
\end{equation}
Clearly, we have $z - \sum_{(i,j)\in C} w_{ij}\geq 0$. 
Initially, we consider the cases where $z - \sum_{(i,j)\in C} w_{ij}= 0$. 
Since $z = \sum_{(i,j)\in C} w_{ij}$, we can conclude that $\sum_{(i,j)\in A'} w_{ij}\leq\sum_{(i,j)\in C} w_{ij}$.
This observation, together with (\ref{des6}), written for each edge $[i,j]$ such that 
$(i,j)\in C$, yields 
\[\kappa-\sum_{(i,j)\in A'} w_{ij}\geq\kappa-\sum_{(i,j)\in C} w_{ij}\geq\sum_{(i,j)\in C} w_{ji}.\]
Now we consider the cases where $z - \sum_{(i,j)\in C} w_{ij}= 1$. The inequality in (\ref{des2:cicloArcos})
is trivially satisfied if $\sum_{(i,j)\in C} w_{ji}\leq\lfloor{\kappa/2}\rfloor$. Let us 
assume that $\sum_{(i,j)\in C} w_{ji}\geq\lfloor{\kappa/2}\rfloor + 1$. Thus, from 
(\ref{des6}) and (\ref{des8}) we can conclude that
\begin{equation}
z\geq\lfloor{\kappa/2}\rfloor + 1
\label{des3:cicloArcos}
\end{equation}
and 
\begin{equation}
\sum_{(i,j)\in C} w_{ij}\leq\kappa-\sum_{(i,j)\in C} w_{ji}\leq\lfloor{\kappa/2}\rfloor.
\label{des4:cicloArcos}
\end{equation}
Notice that the only way to have $z - \sum_{(i,j)\in C} w_{ij}= 1$ is to have $\kappa$ odd and also
(\ref{des3:cicloArcos}) and (\ref{des4:cicloArcos}) 
satisfied with equality. In this case, it follows from satisfying (\ref{des3:cicloArcos}) with equality that
\begin{equation}
\sum_{(i,j)\in C} w_{ji}=z=\lfloor{\kappa/2}\rfloor+1.
\label{des5:cicloArcos}
\end{equation}
From (\ref{des5:cicloArcos}) we can conclude that $\sum_{(i,j)\in A'} w_{ij}\leq\sum_{(i,j)\in C} w_{ji}$, 
and from this we arrive at the desired result, since 
\[\kappa-\sum_{(i,j)\in A'} w_{ij}\geq\kappa-\lfloor{\kappa/2}\rfloor-1=\lfloor{\kappa/2}\rfloor.\]
Finally, in the cases where $z - \sum_{(i,j)\in C} w_{ij}\geq 2$, the inequality in (\ref{des2:cicloArcos}) is
trivially satisfied, since $\sum_{(i,j)\in C} w_{ji}\leq\kappa-1$.
\end{proof}

Notice that the faces defined by the inequalities introduced in 
Theorems~\ref{teo:caminhoK-1}, \ref{kappa:caminhoK-2}, and \ref{teo:cicloArcos} are not 
facets of $P_{G,\kappa}$, since their solutions belong to some face defined by 
(\ref{des6}). Nevertheless, those inequalities can be computationally useful, 
since their solutions are feasible solutions to $AO(G,\kappa)$.

Our last theorem establishes that a structure of $G$ constructed from two paths of length $\kappa$ 
and having some common arcs can induce a valid inequality for $P_{G,\kappa}$.

\begin{thrm}\emph{[Adjacent paths inequality]}
Let $p^\mathrm{I}$ and $p^\mathrm{II}$ be paths in $D$ with $\kappa$ arcs.
Let $\{v^\mathrm{I}_1,\ldots,v^\mathrm{I}_{\kappa+1}\}$ and
$\{v^\mathrm{II}_1,\ldots,v^\mathrm{II}_{\kappa+1}\}$ be the respective vertex sets.
Suppose $p^\mathrm{I}$ and $p^\mathrm{II}$ are such that
\begin{itemize}
\item [(i)]  $v^\mathrm{I}_k=v^\mathrm{II}_k$ for $1\leq k\leq l\leq \kappa$, $l>1$,
\item [(ii)] there exists an edge $[v^\mathrm{I}_r,v^\mathrm{II}_r]\in E$ with $l+1\leq r\leq \kappa+1$. 
\end{itemize}
Then the inequality 
\[w_{v^\mathrm{I}_1v^\mathrm{I}_2} + \sum_{2\leq k < l} 2w_{v^\mathrm{I}_kv^\mathrm{I}_{k+1}} + 
\sum_{l\leq k \leq \kappa} (w_{v^\mathrm{I}_kv^\mathrm{I}_{k+1}} + 
w_{v^\mathrm{II}_kv^\mathrm{II}_{k+1}}) + w_{v^\mathrm{I}_rv^\mathrm{II}_r} + w_{v^\mathrm{II}_rv^\mathrm{I}_r} \leq 2z\]
is valid for $P_{G,\kappa}$.
\label{adjacent paths inequality}
\end{thrm}

\begin{proof}
Consider $(\bar{w},\bar{z})\in P_{G,\kappa}$. Let $d+f$ represent the number of arcs in the set
$p^\mathrm{I}\cup p^\mathrm{II}\setminus \{(v^\mathrm{I}_1,v^\mathrm{I}_2)\}$ having null entries 
in vector $\bar{w}$, with $d$ of them belonging to $p^\mathrm{I}\cap p^\mathrm{II}$ and $f$ belonging either to 
$p^\mathrm{I}$ or to $p^\mathrm{II}$.
When applied to $(\bar{w},\bar{z})$, the adjacent paths inequality imposes a lower bound on $\bar{z}$:
\[\bar{z}\geq \kappa -d -f/2 - (2-\bar{w}_{v^\mathrm{I}_1v^\mathrm{I}_2}-\bar{w}_{v^\mathrm{I}_rv^\mathrm{II}_r}-\bar{w}_{v^\mathrm{II}_rv^\mathrm{I}_r})/2.\]
Next we exhibit a path with $\kappa$ arcs in 
$p^\mathrm{I}\cup p^\mathrm{II}$ which proves this lower bound when the inequality is applied to it.
The following possibilities can happen:

\noindent {(a) $f=0$:} For $p^\mathrm{I}$ we obtain 
\[\bar{z}\geq \kappa - d \geq \kappa -d - (2-\bar{w}_{v^\mathrm{I}_1v^\mathrm{I}_2}-\bar{w}_{v^\mathrm{I}_rv^\mathrm{II}_r}-\bar{w}_{v^\mathrm{II}_rv^\mathrm{I}_r})/2.\]\newline
\noindent {(b) $f>0$:} Consider the case in which $\lfloor f/2\rfloor$ arcs have null entries in a path,
without loss of generality let us say in $p^\mathrm{I}$, and $\lceil f/2\rceil$ arcs have null
entries in the other path, $p^\mathrm{II}$.
For $p^\mathrm{I}$ we get
\begin{equation}
\bar{z}\geq \kappa - d - \lfloor f/2\rfloor,
\label{f>0}
\end{equation}
and, since $\lfloor f/2\rfloor \leq f/2 + (2-\bar{w}_{v^\mathrm{I}_1v^\mathrm{I}_2}-\bar{w}_{v^\mathrm{I}_rv^\mathrm{II}_r}-\bar{w}_{v^\mathrm{II}_rv^\mathrm{I}_r})/2$,
we arrive at the desired result. It is not difficult to verify that any other distribution
of the $f$ arcs between $p^\mathrm{I}$ and $p^\mathrm{II}$ will give us a larger right-hand side of
(\ref{f>0}).
\end{proof}

To finalize, we mention that a result similar to Theorem~\ref{adjacent paths inequality} also holds
if the arcs considered in the statement of the theorem are oriented in the opposite direction.

\section{Concluding remarks}
\label{sec:directions}

We have introduced a formulation of acyclic orientations with path constraints. Our formulation
is related to a more general formulation with diameter constraints and is, therefore, also related
to several combinatorial optimization problems that seek optima on the set of acyclic orientations. We also
presented a partial study of the polytope associated with the formulation, 
introducing further valid inequalities as well. 

The results we have presented open up several possibilities for continued research. One of them is 
to develop separation algorithms that can be used efficiently in a cutting-plane framework.  
Once results are obtained on this front, a first candidate for a study on applications
seems to be the FAP instances with channel separation 1, 2, or 3, 
as discussed in Section~\ref{sec:combProblems}. 
In order to handle the enlarged FAP instances, we first need to develop some preprocessing 
techniques, such as in \cite{ardaal95}.

\subsection*{Acknowledgments}

The authors acknowledge partial support from CNPq, CAPES, and a FAPERJ BBP
grant.

\bibliography{paper}

\begin{thebibliography}{10}

\bibitem{ardaal95}
K.~Aardal, A.~Hipolito, C.~van Hoesel, B.~Jansen, C.~Roos, and T.~Terlaky.
\newblock {EUCLID CALMA} radio link frequency assignment project: A
  branch-and-cut algorithm for the frequency assignment problem.
\newblock Technical report, Delft and Eindhoven Universities of Technology, The
  Netherlands, 1995.

\bibitem{Bermond99}
J.~Bermond, J.~Bond, C.~Martin, A.~Pekec, and F.~Roberts.
\newblock Optimal orientations of annular networks.
\newblock {\em {Journal of Interconnection Networks}}, 1:21--46, 2000.

\bibitem{bermond97acyclic}
J.~Bermond, M.~Di~Ianni, M.~Flammini, and S.~Perennes.
\newblock Acyclic orientations for deadlock prevention in interconnection
  networks.
\newblock In {\em Proceedings of the Workshop on Graph-Theoretic Concepts in
  Computer Science}, pages 52--64, 1997.

\bibitem{Borndorfer98}
R.~{Bornd\"orfer}, A.~{Eisenbl\"atter}, M.~{Gr\"otschel}, and A.~Martin.
\newblock The orientation model for frequency assignment problems.
\newblock Technical Report 98-01, Zuse Institute Berlin, Germany, 1998.

\bibitem{Deming79}
R.W. Deming.
\newblock Acyclic orientations of a graph and chromatic and independence
  numbers.
\newblock {\em {Journal of Combinatorial Theory, Series B}}, 26:101--110, 1979.

\bibitem{Gallai68}
T.~Gallai.
\newblock On directed paths and circuits.
\newblock In P.~Erd\H{o}s and G.~Katona, editors, {\em Theory of Graphs}, pages
  115--118. Academic Press, New York, NY, 1968.

\bibitem{Junger85B}
M.~{Gr\"otschel}, M.~{J\"unger}, and G.~Reinelt.
\newblock Facets of the linear ordering polytope.
\newblock {\em {Mathematical Programming}}, 33:43--60, 1985.

\bibitem{Junger85}
M.~{Gr\"otschel}, M.~{J\"unger}, and G.~Reinelt.
\newblock On the acyclic subgraph polytope.
\newblock {\em {Mathematical Programming}}, 33:28--42, 1985.

\bibitem{Maniezzo2000}
V.~Maniezzo and A.~Carbonaro.
\newblock An ants heuristic for the frequency assignment problem.
\newblock {\em {Future Generation Computer Systems}}, 16:927--935, 2000.

\bibitem{Roy67}
B.~Roy.
\newblock Nombre chromatique et plus longs chemins d'un graphe.
\newblock {\em {Revue AFIRO}}, 1:127--132, 1967.

\end{thebibliography}
\bibliographystyle{plain}

\end{document}